\documentclass[12pt]{iopart}

\usepackage{graphicx}
\usepackage{epsfig,latexsym}

\newcommand{\be}{\begin{equation}}
\newcommand{\ee}{\end{equation}}
\newcommand{\jpsi}{J / \psi}

\begin{document}

\title{Predictions for Quarkonia Dissociation}

\author{\'Agnes M\'ocsy$^1$ and P\'eter Petreczky$^2$}

\address{$^1$ RIKEN-BNL
  Research Center, Brookhaven National Laboratory, Upton NY 11973, USA}
  
 \address{$^2$ Physics Department, Brookhaven National Laboratory, Upton NY 11973, USA} 

\ead{mocsy@bnl.gov}

\begin{abstract}
We predict the upper bound on the dissociation temperatures of  different
quarkonium states.
\end{abstract}


In a recent paper \cite{mocsy07-1} we analyzed in detail the quarkonium spectral functions. This analysis has shown that spectral functions
calculated using potential model for the non-relativistic Green's function combined with perturbative QCD can describe the available lattice data on quarkonium correlators both at zero and finite temperature in QCD with no light quarks \cite{mocsy07-1}. Charmonia, however, were found to be dissolved  at temperatures significantly lower than quoted in lattice QCD studies, and in contradiction with other claims made in recent years from different potential model studies. 
In \cite{mocsy07-2} we extended the analysis to real QCD
with one strange quark and two light quarks using new lattice QCD data on quark anti-.quark free
energy obtained with small quark masses \cite{kostya}.  

Here we briefly outline the main results of the analysis of \cite{mocsy07-2}, in particular the estimate for the upper limit on the dissociation temperatures. 
There is an uncertainty 
in choosing the quark-antiquark potential at finite temperature. In \cite{mocsy07-2} we considered two
choices of the potential, both consistent with the lattice data \cite{kostya}.  
The more extreme choice, still compatible with lattice
data, leads to the largest possible binding energy.  In this most binding potential some of the quarkonium states survive above deconfinement, but their strongly temperature-dependent binding energy is
significantly reduced. This is shown in Fig.~\ref{fig:binding}. 
 Due to the  reduced binding energy thermal activation
can lead to the dissociation of quarkonia, even when the corresponding peak is
present in the spectral function. Knowing the binding energy we estimate the thermal width
using the analysis of  \cite{Kharzeev:1995ju}. The expression of the rate of thermal excitation has
particularly simple form in the two limiting cases:
\begin{figure}
\begin{center}
\includegraphics[width=6.5cm]{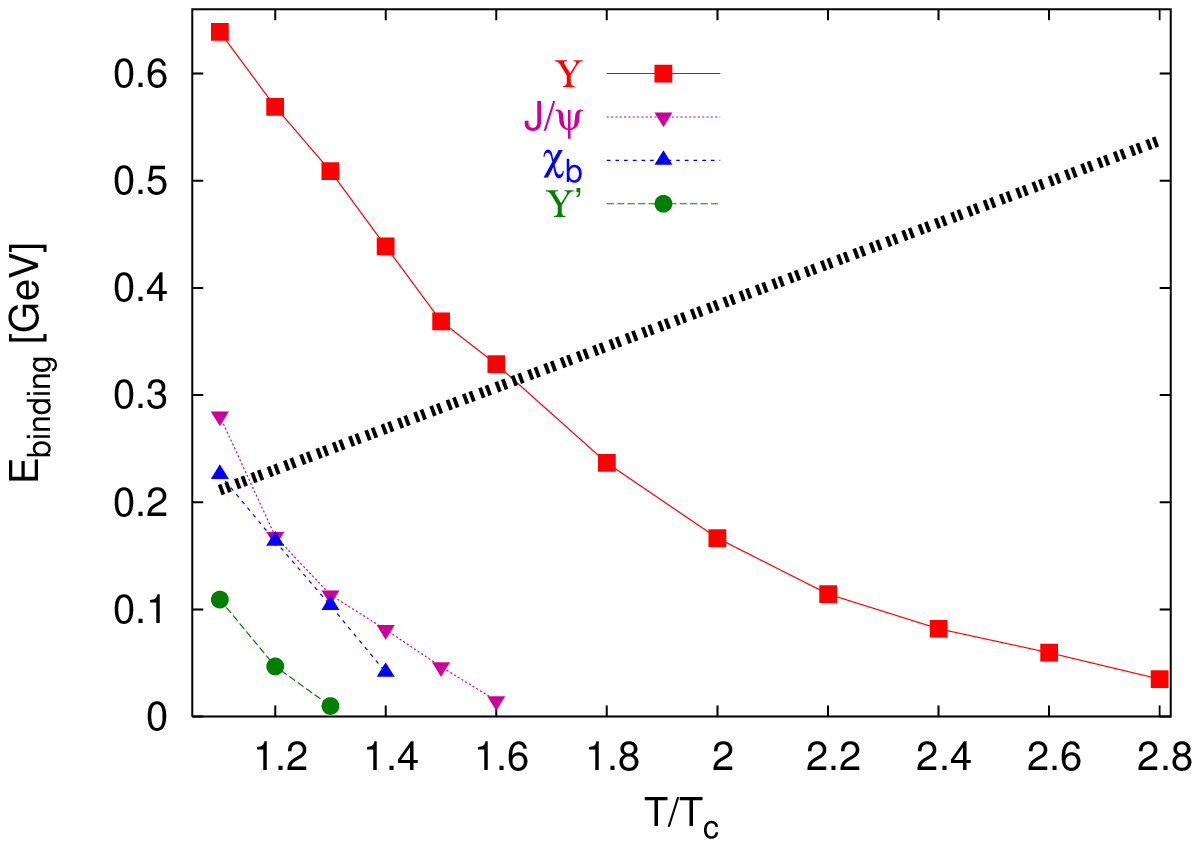}
\includegraphics[width=6.5cm]{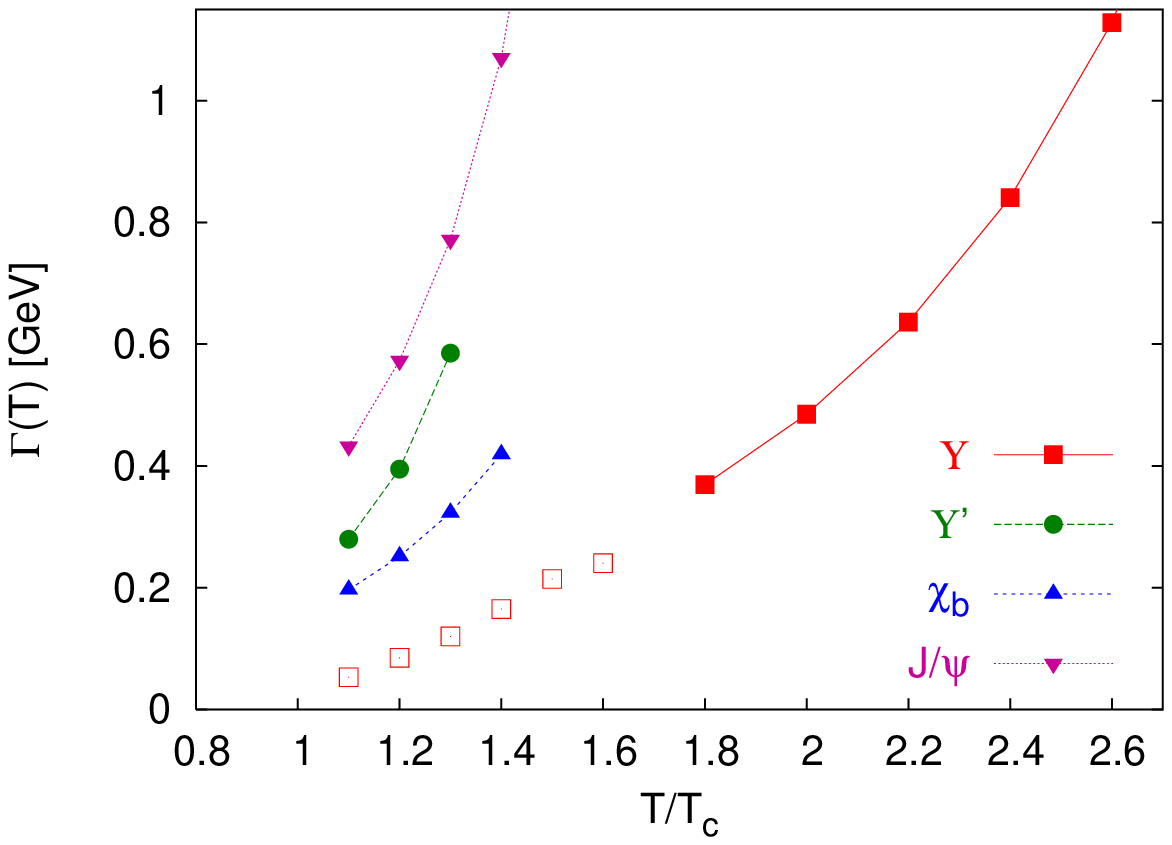}
\caption{Upper limit of the binding energy (left) and the width (right) of quarkonium states. For better visibility, in the limit of small binding, the open squares show the width of the 1S bottomonium state multiplied by six.}
\label{fig:binding}
\end{center}
\end{figure}
\be
\Gamma(T)=\frac{(LT)^2}{3\pi}Me^{-E_{bin}/T}\, ,~E_{bin}\gg T\, \quad 
\Gamma(T)=\frac{4}{L}\sqrt{\frac{T}{2\pi M}}\, ,~ E_{bin}\ll T \, .
\label{small}
\ee
Here $M$ is the quarkonium mass, $L$ is the size of the spatial region of the potential, given by the distance from the average 
quarkonium radius to the top of the potential, i.e. $L=r_{med}-\langle r^2\rangle^{1/2}$, $r_{med}$ 
being the effective range of the potential \cite{mocsy07-2}. 
Using the above formulas we estimate the thermal width of charmonium and bottomonium states.
Since in the deconfined phase $E_{bin}<T$ the $1S$ charmonium and $2S$ and $1P$ bottomonium states are in the regime of weak
binding, and their width is large, as shown in Fig.~\ref{fig:binding}. The $1S$ bottomonium is strongly bound for $T<1.6T_c$ and its thermal
 width is smaller than $40$MeV. For $T>1.6T_c~$, however, even the 1S bottomonium states
 is in the weak binding regime resulting in the large increase of the width, see Fig.~\ref{fig:binding}.
 When the thermal
 width is significantly larger than the binding energy no peak structure will be present in the
 spectral functions, even though the simple potential model calculation predicts a peak.
 Therefore, we define a conservative dissociation temperature by the condition
 $\Gamma>2 E_{bin}$. The obtained dissociation temperatures are
 summarized in Table \ref{tab:diss}. 
  
From the table it is clear that all quarkonium states, except the $1S$ bottomonium, will melt at temperatures considerably smaller than previous estimates, and will for certain be dissolved in the matter produced in heavy ion collision at 
LHC. Furthermore, it is likely that at energy densities reached at the LHC a large fraction of the $1S$ bottomonium states will also dissolve. It has to be seen to what extent these findings will result in large $R_{AA}$ suppression at LHC. For this more information about initial state effects is needed. Moreover, the spectral
functions are strongly enhanced over the free case even when quarkonium states are
dissolved \cite{mocsy07-1,mocsy07-2} indicating significant correlations between the heavy quark and antiquark. Therefore, one should take into account also the possibility of quarkonium regeneration from correlated initial quark-antiquark pairs.

\begin{table}[h]
\renewcommand{\arraystretch}{0.81}
\begin{center}
\begin{minipage}{8.5cm} \tabcolsep 5pt
\begin{tabular}{|ccccccc|}
 state&$\chi_c$&$\psi'$&$\jpsi$&$\Upsilon'$&$\chi_b$&$\Upsilon$\\ \hline 
 $T_{dis}$&$\le T_c$& $\le T_c$ &$1.2T_c$&$1.2T_c$&$1.3T_c$&$2T_c$\\ 
\end{tabular}
\end{minipage}
\caption{Upper bound on quarkonium dissociation temperatures.}
\end{center}
\label{tab:diss}
\end{table}

\end{document}